\documentstyle[11pt,fewbody2005,twoside,epsfig]{article}

\newcommand{\be}{\begin{equation}} 
\newcommand{\ee}{\end{equation}} 
\newcommand{\bea}{\begin{eqnarray}} 
\newcommand{\eea}{\end{eqnarray}}  
\newcommand{\bean}{\begin{eqnarray*}} 
\newcommand{\eean}{\end{eqnarray*}}

\def\lsim{\raise 0.4ex\hbox{$<$}\kern -0.8em\lower 0.62ex\hbox{$\sim$}} 
\def\gsim{\raise 0.4ex\hbox{$>$}\kern -0.7em\lower 0.62ex\hbox{$\sim$}}

\newcommand{\bse}{\begin{subequations}}
\newcommand{\ese}{\end{subequations}}

\newcommand{\cd}{{\langle n(r) \rangle_p}}

\def\spose#1{\hbox to 0pt{#1\hss}}      
\def\ltapprox{\mathrel{\spose{\lower 3pt\hbox{$\mathchar"218$}}      
\raise 2.0pt\hbox{$\mathchar"13C$}}}      
\def\gtapprox{\mathrel{\spose{\lower 3pt\hbox{$\mathchar"218$}}      
\raise 2.0pt\hbox{$\mathchar"13E$}}}      
\def\inapprox{\mathrel{\spose{\lower 3pt\hbox{$\mathchar"218$}}      
\raise 2.0pt\hbox{$\mathchar"232$}}}

\begin{document}

\title{Gravitational  Structure Formation}

\author{Francesco Sylos Labini}
\affil{``E. Fermi'' Center, Via Panisperna 89 A, Compendio del 
Viminale, 00184 - Rome, Italy and ISC-CNR Via dei Taurini, 19 00185
Rome, Italy}

\author{Thierry Baertschiger}
\affil{Dipartimento di Fisica, 
Universit\`a ``La Sapienza'', P.le A. Moro 2, I-00185 Rome, Italy and
ISC-CNR Via dei Taurini, 19 00185 Rome, Italy}

\begin{abstract}
We discuss the formation of the first structures in gravitational
N-body simulations.  The role of two-body interaction is found to be a
crucial element and an analogy with the dynamics of the Coulomb
lattice, well-studied in solid state physics, is discussed.
\end{abstract}
\keywords{Gravitational many body simulations}

The standard models of the formation of large scale structure of the
universe is based on the gravitational growth of small initial density
fluctuations in a homogeneous and isotropic medium (e.g. Peebles
1980). For example, in the so-called Cold Dark Matter (CDM) model,
particles interact only gravitationally and they are cold, i.e. with
very small initial velocity dispersion. This situation allows one to
model this system with a collision-less Boltzmann equation and, for
sufficiently large scales, pressure-less fluid equations. Then it is
possible to solve in a perturbative way, for small density
fluctuations, these fluid equations (see e.g. Peebles 1980). However
this treatment is inapplicable in the strong non-linear regime. Then,
the most widely used tool to study gravitational clustering in the
various regimes is by means of N-body simulations (NBS) which are
based on the computation of the dynamics of self-gravitating particles
in expanding universe.

These simulations can be performed by considering an infinite periodic
system, i.e. a finite system with periodic boundary condition. Despite
the simplicity of the system, in which dynamics are Newtonian at all
but the smallest scales, the analytic understanding of this crucial
problem is limited to the regime of very small fluctuations where a
linear analysis can be performed. In the cosmological case, the
problem can be approximated to Newtonian but the equation of motions
are modified because of the expanding background (Peebles, 1980).  As
discussed below, we find instructive to consider some simplified cases
where the expansion is not included and then study the differences
introduced by space expansion.

An important point should be stressed: for numerical reasons, the
cosmological density field must be discretized into
``macro-particles'' interacting gravitationally which are tens of
order of magnitudes heavier than the (elementary) CDM particles due to
computer limitations. This procedure introduces discreteness at a much
larger scale than the discreteness inherent to the CDM particles. By
discreteness we mean statistical and dynamical effects which are not
described by the self-gravitating fluid approximation. The
discreteness has different manifestations in the evolution of the
system (see e.g. Baertschiger et al., 2002 and references therein).
In this context it is necessary to consider the issue of the physical
role of discrete fluctuations in the dynamics, which go beyond a
description where particles play the role of collision-less fluid
elements and the evolution can be understood in terms of a
self-gravitating fluid.

\begin{figure} [h]
\begin{center}
\epsfig{file=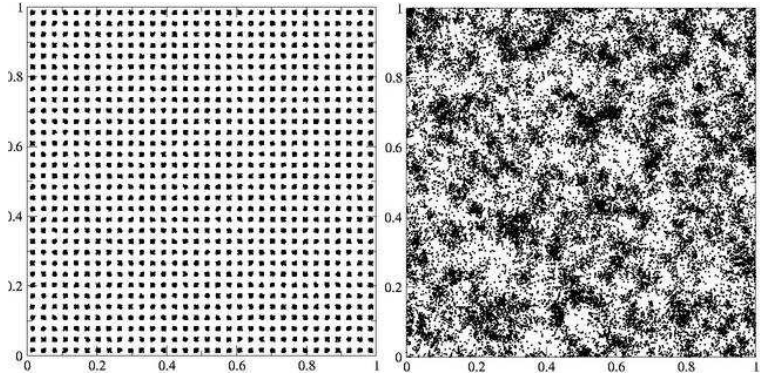,width=100mm,angle=0}
\caption {{\it Left Panel}: 
Slightly perturbed lattice with zero initial velocity dispersion. This
is an orthogonal projection of a system with $32^3$ particles.  The
force on a particle is small and linearly proportional to the
displacements of all the particles from their lattice positions. {\it
Right Panel}: When the system is evolved under its own gravity for
long times it creates complex non-linear structures characterized by
the presence of clusters of different sizes. When the size of the
largest cluster becomes of the order of the box size, the simulation
is dominated by finite size effects. }
\label{lattice1}
\end{center}
\end{figure}

In order to study the gravitational many-body problem, we have
considered a paradigmatic system consisting of a very simple initial
particle distribution represented by a slightly perturbed simple cubic
lattice with periodic boundary conditions and zero initial velocities
(Joyce et al., 2005). A perfect cubic lattice is an unstable
equilibrium configuration for gravitational dynamics, being the force
on each particle equals zero. A slightly perturbed lattice (see
Fig.\ref{lattice1}) represents instead a situation where the force on
a particle is small and linearly proportional to the displacements of
all the particles from their lattice positions. When the system is
evolved for long enough times it creates complex non-linear structures
as shown in Fig.\ref{lattice1} (Right Panel). While the full
understanding of this clustering dynamics is not currently available,
some steps have been done for what concerns the early times evolution
of the system (Joyce et al., 2005, Baertschiger \& Sylos Labini,
2004).

\noindent

The characterization of the gravitational evolution of this system for
small displacements, that is up to when two nearest particles collide,
can be achieved by a perturbative theory (Joyce et al., 2005).  
Up to a change in sign in the force, the initial configuration is
identical to the Coulomb lattice (or Wigner crystal) in solid-state
physics (see e.g. Pines, 1963): by using standard techniques of 
solid-state physics it is possible to develop an approximation to the
evolution of the gravitational many-body problem.  The equation of
motion of particles moving under their mutual self-gravity in a static
universe is 
\be
{\ddot {\bf x} }_i 
= - \sum_{i\neq j} 
\frac{G m_j ({\bf x}_{i}-{\bf x}_j) }{|{\bf x}_{i}-{\bf x}_{j}|^3}\,,
\label{eom}
\ee
where for seek of clarity we have not written explicitly the sum on
the replicas of the system.  Here dots denote derivatives with respect
to time $t$, ${\bf x}_i$ is the position of the $i$th particle, of
mass $m_i$ \footnote{Note that as written in Eq.~(\ref{eom}) the
infinite sum giving the force on a particle is not explicitly well
defined. It is calculated by solving the Poisson equation for the
potential, with the mean mass density subtracted in the source
term. In the cosmological case this is appropriate as the effect of
the mean density is absorbed in the Hubble expansion; in the case of
the Coulomb lattice it corresponds to the assumed presence of an
oppositely charged neutralizing background.  In the non-expanding case
the the negative background can be intended as a trick used to make
the potential finite: however in the conditions we consider this does
not affect the computation of the force}.  Perturbations from the
Coulomb lattice are described simply by Eq.~(\ref{eom}), with $Gm^2
\rightarrow -e^2$ (where $e$ is the electronic charge). By denoting
${\bf x}_i(t)={\bf R} + {\bf u}({\bf R},t)$ where ${\bf R}$ is the
lattice vector of the $i$th particle and ${\bf u}({\bf R},t)$ is the
displacement of the particle from {\bf R}, and by expanding to linear
order in ${\bf u}({\bf R},t)$ about the equilibrium lattice
configuration (in which the force on each particle is exactly zero),
we obtain
\be
{\bf {\ddot u}}({\bf R},t) 
= - \sum_{{\bf R}'} 
{\cal D} ({\bf R}- {\bf R}') {\bf u}({\bf R}',t)\,. 
\label{linearised-eom}
\ee
The matrix ${\cal D}$ is called in solid state physics, for any
interaction, the {\it dynamical matrix} which, according to the Bloch
theorem, is diagonalized by plane waves in reciprocal space. The
spectrum of eigenvalues is complex and as in the case of the Coulomb
lattice, eigenvectors present the characteristic branch structure.

For example, in the Coulomb lattice there is the {\it optical} branch,
describing oscillations with plasma frequency $\omega_p^2=4 \pi e^2
n_0/m$ (where $n_0$ is the electronic number density) which in the
gravitational case corresponds, for long wavelength perturbations, to
the evolution of instabilities predicted by an analogous fluid
description of the self-gravitating system
(Joyce et al., 2005). Further it is possible to characterize precisely,
up to when two nearest particles collide, the deviations from this
fluid-like behavior at shorter wavelengths arising from the discrete
nature of the system. For instance there are also oscillating modes,
and modes which grow faster than the fluid one, which are absent in
the fluid description.

This analysis should be a useful step toward a precise quantitative
understanding, which is currently lacking, of the role of discreteness
in cosmological NBS (see e.g.Melott and Shandarin, 1993).  These
simulations are most usually started from configurations which are
simple cubic lattices perturbed in a manner prescribed by a
theoretical cosmological model and thus the early times dynamical
evolution can be studied as the paradigmatic case discussed above,
with only a simple modification of the dynamical equations due to the
expansion of the Universe. The main difference is quantitative, namely
in the expanding case the growth of perturbations is power-law in time
while in the non-expanding case it is exponential (Joyce et al.,
2005). Apart this, no qualitative physical difference in the formed
non-linear structures is apparent (see also Sylos Labini et al., 2004).

One of the central questions in the context of gravitational NBS is
whether one may have some analytical predictions which relate the
observed power-law in the correlation function of the particles at
late times with some features of the initial particle configuration.
For example it has been recently observed~(Sylos Labini et al., 2004)
that in a broad class of gravitational NBS it is shown a universal
behavior in the non-linear clustering which develops, characterized by
the {\it exponent} of the conditional density. This statistical
quantity is defined as
\be
\label{cd} 
\langle n(r) \rangle_p
= \frac{\langle n(r)n(0) \rangle} {\langle n(0) \rangle} \;,
\ee
so that $\left<n(\vec{r})\right>_p dV$ gives the a-priori probability
of finding $1$ particles placed in the infinitesimal volumes $dV$
around $\vec{r}$ with the {\it condition} that the origin of
coordinates is {\it occupied} by a particle, i.e.  it represents the
average density of particles seen by a fixed particle at a distance
$r$ from it.  Once power-law correlations are developed,
i.e. $\cd\approx r^{-\gamma}$ with $\gamma\approx 1.8$, the subsequent
evolution increases the range of scales where non-linear clustering is
formed, i.e.  where $\langle n(r) \rangle_p \gg n_0$, by
approximatively a simple rescaling: denoting by $\langle
n(r,t)\rangle_p$ the conditional density at time $t$, one has
\bea
\langle n(r,t+\delta)\rangle_p \approx \langle n(a\cdot r,t)\rangle_p 
\label{eq:rescaling}
\\
\nonumber 
\cd\approx r^{-\gamma} \;\; \mbox{for}\;\; r < \lambda_0(t)
\eea
where $a<0$ is a constant which depends on the time (Baertschiger and
Sylos Labini, 2004) and $\lambda_0(t)$ is the crossover scale between
strong ($\langle n(r) \rangle_p \gg n_0$) and weak ($\langle n(r)
\rangle_p \approx n_0$) clustering.  While the constant $a$, as well
as $\lambda_0(t)$, depends on the particular system considered it
seems that the exponent $\gamma$ is the same in many different cases
(Sylos Labini et al., 2004). Thus an important element of the nature
of clustering in the non-linear regime can be associated with what is
common to all these different simulations: their evolution in the
non-linear regime is dominated by fluctuations at small scales, which
are similar in all cases at the time this clustering develops. Such
``shot-noise'' fluctuations are in fact intrinsic to any particle
distribution. This corresponds to domination by nearest neighbor
interactions when the first non-linear structures are formed
(Baertschiger and Sylos Labini, 2004).

To study in detail this early non-linear dynamics, i.e. the growth of
the first non-linear correlations, we considered the gravitational
evolution of a cold particle distribution with no correlations, i.e. a
Poisson configuration (Baertschiger and Sylos Labini, 2004). One may
show that by treating this simple case in a static universe as an
ensemble of isolated two-body systems, one may understand the origin
of the first non-linear correlated structures.  This is possible
because: (i) the full gravitational force probability distribution
approaches the nearest-neighbor force probability distribution at
large values of the field and (ii) most of particles are mutually
nearest-neighbors.  The exponent of the conditional density is then
simply related to the functional form of the time evolved nearest
neighbors probability distribution, whose time dependence can be
computed by using Liouville theorem for the gravitational two-body
system (Baertschiger and Sylos Labini, 2004).

The fundamental open problem is that of understanding whether large
non-linear structures, which at late times contain many particles, are
produced solely by collision-less fluid dynamics, or whether the
particle collisional processes (i.e. discreteness effects) are
important also in the long-term, or whether they are made by a mix of
these two effects. Important elements in this respect are represented
by the fact that the correlation function of the evolved system has a
strikingly similar functional form to the one generated at early times
by two-body interaction (described by Eq.\ref{eq:rescaling}) and by
the fact that aggregation proceeds in a hierarchical bottom-up manner
(Baertschiger and Sylos Labini, 2004).

\acknowledgments 
We thank A. Gabrielli, M. Joyce and B. Marcos for fruitful
collaborations and W.C. Saslaw for useful discussions.  We also
acknowledge the ``Centro Ricerche e Studi E. Fermi'' Roma for the use
of a super-computer to make numerical calculations.

\end{document}